# Tempering Dynamics and Relaxation Times in the 3D Ising Model

Luis A. Fernández[(a)], Enzo Marinari[(b)] and Juan J. Ruiz-Lorenzo[(c)]

(a): Departamento de Física Teórica, Universidad Complutense de Madrid

Ciudad Universitaria , Madrid 28040 (Spain)

laf@lattice.fis.ucm.es

(b): Dipartimento di Fisica and Infn, Università di Cagliari

Via Ospedale 72, 09100 Cagliari (Italy)

marinari@ca.infn.it

(c): Dipartimento di Fisica and Infn, Università di Roma La Sapienza

P. A. Moro 2, 00185 Roma (Italy)

ruiz@chimera.roma1.infn.it

December 5, 1994

**Abstract**

We discuss the tempering Monte Carlo method, and its critical slowing down in the 3d Ising model. We show that at $T_c$ the tempering does not change the critical slowing down exponent $z$. We also discuss the exponential slowing down for the transition from the plus to the minus state in the cold phase, and we show that tempering reduces it to a power law slowing down. We discuss the relation of the flip-flop rate to the surface tension for the local dynamical schemes.

1

# 1 Introduction

The tempering updating method [1] has been introduced recently in order to speed up simulations of statistical systems. Methods very similar in nature were already in use in the framework of molecular dynamics simulation [2]. Multicanonical methods [3] also have deep similarities with the tempering approach.

The main idea of tempering (see also [4, 5, 6, 7, 8, 9]) is to let the temperature to become a dynamical variable, and to let it vary in the course of the dynamics. Tempering is in some sense a sophisticated kind of annealing, where the system is always at thermal equilibrium. Already in its first proposal and implementation of ref. [1] the tempering method turned out to be very effective when employed to study the low $T$ phase of the $3d$ Random Field Ising Model.

There are some important advantages in the tempering approach (and in this note we will try to establish a new positive evidence). The first one is maybe that tempering is very simple, and that the spin configurations one generates by the tempering dynamics are distributed according to the Boltzmann distribution. There is no need for any weighting or reanalysis (improved estimation schemes can obviously be employed). We want also to stress that tempering parallelization is trivial. The algorithm is local, but for the use of a rare global communication, in the form of an add-reduce function.

Apart from the technical advantages we have just recalled there are important physical reasons for which methods like tempering are becoming important. Indeed an improved Monte Carlo approach seems crucial, for example, for an effective numerical study of finite dimensional spin glasses [4, 8, 10, 6, 7]. The problem of the behavior of three dimensional system is still quite open, and improved numerical methods look here crucial. The main issue is the transition among different local minima of the free energies. In the course of the dynamics the system gets trapped in local minima, and the warming up allowed by the tempering favors the exploration of different local minima.

Here we study the normal, non disordered, $3d$ Ising model. Here in the cold phase there are two states, and we will look at the transition rate among these two states as prototype for a general transition among local minima. The Ising model is interesting since it is the simplest possible model, and since in this case it is very clear what going from a given state to another one means (one uses the magnetization to monitor the state to state transition). We will show that tempering allows to reduce the exponential slowing down that local methods like heat bath updating encounter to a power law slowing down.

In Section (2) we define the quantities that we will discuss in the following. In Section (3) we discuss the autocorrelation time at the critical point $T = T_c$. We show that, as we would have expected, in this case tempering cannot change the critical exponent $z$. In Section (4) we discuss two main issues. In first we show that by counting flip-flops of a local dynamics we can get hints about the surface tension. In second we show that if we let the tempering wander (in $\beta$ space) only in the cold phase the exponential slowing down survives. In Section (5) we show that if we let the tempering pass the border of $T_c$ and change $T$ among the two phases the slowing down becomes power law like. This is the main result of this note, and we shortly describe its main implications is Section (6).

## 2 A Few Definitions

In the following we will give a few definitions concerning typical time scales of the random dynamics underlying our evaluation of the Ising model partition function. So doing we will be mainly following Alan Sokal very comprehensive book-like lecture notes [11].

Let us start by stressing that correlation times can be defined for the different observables ($A$, let us say) of the theory, and generically they will not need to be equal. We define the *exponential correlation time* by the large time decay of the connected correlation function

$$\rho_A(t) \equiv \langle A(0)A(t)\rangle_c \simeq C \exp\{-\frac{t}{\tau_A^{(exp)}}\} \;, \tag{1}$$

for $t \to \infty$, where $A(t)$ is the value that the observable quantity $A$ takes at the time $t$, and we take the connected part (indicated by the subscript $c$) by subtracting[1] $\langle A \rangle^2$. As we have explained we are explicitly remarking the $A$ dependence of this time scale, and by the suffix ($exp$) the fact that it has been defined from the exponential decay of the correlation function. This asymptotic behavior of the correlation function is corrected, at finite time, by sub-leading contributions. We can have, for example, logarithmic corrections in the exponent, and a sum of a series of time dependent contributions to the correlation function.

Integrating $\rho(t)$ over all times $t$ we can define the *integrated correlation time*

$$\tau_A^{(int)} \equiv \frac{1}{2\rho(0)} \sum_{t=-\infty}^{\infty} \rho(t) \;, \tag{2}$$

where we have assumed a discrete time dynamics (the extension to a continuum time dynamics is straightforward), and the factor 2 makes the integrated time equal to the exponential time in the case of a pure exponential decay (at all times) of the correlation function.

The integrated correlation time has a special relevance, since it is strictly connected to the statistical error which affects the numerical data. Indeed, again following [11], one can relate directly $\tau_A^{(int)}$ to the true statistical error over $A$. Let us define $\sigma_A^{(NAIVE)}$ as the naive fluctuations of $A$. For a dynamics of $T$ steps (starting after thermalization, *i.e.* already at thermal equilibrium) if $\langle A \rangle$ is the expectation value of $A$ we define

$$\sigma_A^{(NAIVE)} \equiv \sqrt{\frac{\sum_t (A_t - \langle A \rangle)^2}{T(T-1)}} \;, \tag{3}$$

---

[1] We are assuming we have reached thermal equilibrium, and our time series is invariant under time translations.

that would be the true error over $A$ if the configurations collected during the dynamics were uncorrelated. On the contrary we call $\sigma_A^{(TRUE)}$ the true statistical error over $A$. $\sigma_A^{(TRUE)}$ can be estimated for example by binning the measured quantities $A_t$ in bins large enough to make the $\sigma$ estimated independent from the bin size. One finds that

$$\tau_A^{(int)} = \left(\frac{\sigma_A^{(TRUE)}}{\sigma_A^{(NAIVE)}}\right)^2 \ . \tag{4}$$

One can roughly estimate that, as far as the value of the observable $A$ is concerned, the number of independent configurations produced by the algorithm is $\frac{T}{2\tau_A^{(int)}}$.

At a critical point, where a correlation length $\xi$ is diverging when $T \to T_c$, the correlation times do typically diverge. One gets that the integrated correlation time

$$\tau_A^{(int)} \simeq \xi^{z_A^{(int)}} \ . \tag{5}$$

The usual argument tells that a local dynamics has $z \geq 2$, since the system is undergoing a random walk in configuration space. Transmitting information from site $x$ to site $y$ at distance $\delta$ with a pure random walk takes a time of order $\delta^2$. Since we need to carry information at a distance of order $\xi$, 2 turns out to be a lower bound for $z$ (additional slowing down phenomena can make $z$ larger).

In order to study the tunneling relaxational dynamics we define a tunneling time $\tau_T$ by counting the number of steps needed to the system to go from the (let us say) plus ground state to the minus ground state and back. On a finite lattice this definition can be plagued from ambiguities, since it is not completely clear when the system has completed a transition. That defines a $\beta$ window in which for a given lattice size things go smoothly. We are in the broken phase. $\beta$ cannot be too close to $\beta_c$ since in this case the transition signature becomes murky. But $\beta$ cannot be too high or we never get transitions (which we call *flip-flop*), at least when using the normal local Monte Carlo dynamics. In this paper we have tried to work with $\beta$ values which satisfy this constraint. We did not seem to have problems in defining in a non-ambiguous way the transition from a ground state to the other one.

When using the tempering all correlation times are defined by first selecting the configurations which were characterized by the relevant $\beta$ value. That means that the real computer time taken for a given tempered measurement is $N_\beta$ times larger than the one we quote here. The scaling of $N_\beta$ has always to be accounted for when analyzing the scaling properties of the correlation times.

## 3 The Autocorrelation Time at $T = T_c$

In this section we will discuss the behavior of the correlation time at $T = T_c$. We will compare the usual local heat bath dynamics (HB) with the Swendsen-Wang (SW) algorithm (which severely reduces critical slowing down). We will apply tempering to both approaches (obtaining HBT and SWT, with a quite obvious notation). We will show numerically that in both cases tempering does not change the critical exponent $z$.

It is easy to develop an intuitive argument[2] suggesting that tempering cannot eliminate the critical divergence of the correlation time which is proper of the underlying algorithm (for example Heat Bath dynamics). It is clear indeed that if we want the critical slowing down to be defeated we have to let $\beta$ variate in the range $[\beta_m(L), \beta_M(L)]$ such that the correlation length at the two extremes is fixed in *lattice* units. But in order to do that in the infinite volume limit we will need a divergent number of $\beta_\alpha$ values around $\beta_c$ (since the $\delta\beta$ allowed to keep fixed acceptance of the $\beta$ change will be asymptotically infinitesimal).

In slightly more quantitative terms one can start from the fact that the optimal $\delta\beta$ (which guarantees, let us say, an acceptance factor of the order of 50%) is [1, 4, 8]

$$\delta\beta^2 \simeq \frac{1}{\langle H^2 \rangle - \langle H \rangle^2} = \frac{1}{V C_V}, \qquad (6)$$

*i.e.* is connected to the fluctuations of the internal energy of the system, and to the specific heat of the system. Using the fact that at the critical point the specific heat scales as

$$C_V \simeq L^{\frac{\alpha}{\nu}} \qquad (7)$$

we get that the optimal value of $\delta\beta$ scales as $L^{-\frac{1}{2}(d+\frac{\alpha}{\nu})}$ (or using the scaling relation as $L^{-\frac{1}{\nu}}$), and goes to zero in the infinite volume limit. Notice that at $\beta_c \pm \delta\beta$ the correlation length scales as $L$ and consequently with a fixed number of $\beta$ values the $z$ exponent should not change. If we try to keep the system at a fixed correlation length for some $\beta$ the number of $\beta$ values needed diverges.

In order to check this scenario we have simulated the $3D$ Ising model by using the Heat Bath and the Swendsen Wang algorithm. To both of them we eventually added a tempering part (and as we already said we denote the two modified algorithms by HBT and SWT respectively). In all cases we have allowed 5 $\beta$ values to the tempering procedure. The central $\beta$ value has been kept fixed to $\beta_c \simeq 0.22165$, and the $\delta\beta$ has been changed when changing the lattice size. We have used $\delta\beta = 0.005158$ for $L = 16$, $\delta\beta = 0.001642$ for $L = 32$ and $\delta\beta = 0.000527$ for $L = 64$. So $\beta_m(L) = \beta_c - 2\delta\beta(L)$, and $\beta_M(L) = \beta_c + 2\delta\beta(L)$.

To estimate the autocorrelation times and its error bars we have used a self-consistent truncation window algorithm with width 6 $\tau_{\text{int},A}$ as described in the references [12, 11]. We give the number of full lattice sweeps and our estimates for the integrated correlation time for SW and SWT in Table 1. We have estimated the integrated correlation time for different observable quantities (the energy, the absolute value of the magnetization and the magnetization squared).

We have analyzed the data of table 1 using the form (5). Our best estimates for the $z$ values are given in table 2. For completeness we add the Wolff results obtained for the $3D$ Ising model with the Swendsen-Wang algorithm [13, 14].

---

[2] We thank Sergio Caracciolo for a discussion of this point.

| L  | Sweeps | $\tau_E$ | $\tau_{|M|}$ | $\tau_{M^2}$ |
|----|--------|----------|--------------|--------------|
| Swendsen-Wang | | | | |
| 16 | 4.7    | 5.40(9)  | 4.87(8)      | 5.27(9)      |
| 32 | 4.7    | 7.94(16) | 6.9(1)       | 7.6(1)       |
| 64 | 4.8    | 11.3(3)  | 9.6(2)       | 10.7(2)      |
| Swendsen-Wang with Tempering | | | | |
| 16 | 13.9   | 5.7(1)   | 5.04(8)      | 5.47(9)      |
| 32 | 14.7   | 7.91(18) | 6.91(14)     | 7.50(16)     |
| 64 | 8.9    | 10.8(3)  | 9.1(3)       | 10.0(3)      |

Table 1: The number of full lattice sweeps (in unit of $10^6$), and our best estimates for the autocorrelation times. All runs have been started at equilibrium, after a number of thermalization sweeps.

| Method    | $z_E$   | $z_{|M|}$ | $z_{M^2}$ |
|-----------|---------|-----------|-----------|
| SW(Wolff) | 0.50(3) |           | 0.50(3)   |
| SW        | 0.54(2) | 0.49(2)   | 0.51(2)   |
| SW & T    | 0.46(2) | 0.43(2)   | 0.44(2)   |

Table 2: Our results for the critical dynamical exponents for SW and SWT. We also report the data obtained by Wolff with the Swendsen Wang algorithm (SW(Wolff)).

Our results for SW are compatible with the Wolff results (our statistical sample is 4 times larger than the former one). The results we obtain for SWT are lower than the SW ones, but of a very small amount, already quite compatible if we only consider the statistical error. If we also consider the large (unknown) systematic error (we are asking an asymptotic, correction free value from three not so huge lattice sizes) we can safely state that the two sets of results are fully compatible.

The HB and HBT data are also completely compatible with what we expected. In the local dynamics $z \simeq 2$ in all cases, and the tempering only changes the non-universal constant factor.

So, tempering does not change the divergence rate of the correlation times at the critical point. This was to be expected, since the tempering is not really changing the critical dynamics of the system, but is favoring the jumps from one sector of the broken phase to an other one. Next we will study the correlation time in the broken phase, and will see that here the tempering can be a crucial help.

# 4 Below $T_c$: the Relaxation Time when Tempering in the Wrong Phase

In this section we will obtain two results. One will be based on the HB method. We will show that by measuring directly the tunneling time by heat bath we get an estimate of a quantity that thanks to the computation of ref. [15] is indeed connected to the surface tension (even if the quantitative agreement is not perfect). The second result will be negative (as expected), and useful to prepare the positive result of the following section. We will show that when running the tempering on a set of $\beta$ values which remain in the cold phase and do not enter the warm phase (*i.e.* $\beta_m > \beta_c$) the tempered Heat Bath algorithm undergoes the usual exponential slowing down (with a coefficient very close to the one would expect to get at the highest values of $T$ included in the values selected for the tempering scheme).

In the low temperature phase the equilibrium distribution between coexistent phases is $\sim \exp(-cL^{d-1})$, since the coexistent phases are separated by free energy barriers of order $L^{d-1}$ (surface free energy). In this case the conventional local algorithms (*e.g.* Heat-Bath or Metropolis) undergo an extremely severe slowing down[3]. The autocorrelation time, tunneling or ergodic time in this context, behaves as

$$\tau \sim \exp(cL^{d-1}) \ . \tag{8}$$

So while at $T_c$ the system undergoes a critical "power" slowing down, for $T < T_c$ an "exponential" slowing down exists, induced by the slow motion from one pure state to the other one.

The previous formula (8), for the $d$-dimensional Ising model, can be deduced through the calculus of the mass gap *i.e.* the inverse of the tunneling time) using semiclassical methods and considering only the contribution of one instanton [15]. One gets in this way an estimate for the constant $c$ of (8). The results is:

$$\tau \sim \exp(2\sigma(T)L^{d-1}) \ . \tag{9}$$

One can identify $\sigma(T)$ with the surface free energy. The two in the argument of the exponential comes from the use of periodic boundary conditions in the semiclassical calculus (when using periodic boundary conditions one has to create two interfaces). This formula (9) is also relevant when discussing the Ising model in a cylindrical geometry [16].

Our results obtained by using the Heat Bath dynamics show that indeed the result (9) can be substantiated numerically. For example at $\beta = 0.223$ and lattice sizes $L = 10, 16, 20$ and 24 we fit our results for $\tau_T$ by

$$\log \tau = a_1 + a_2 L^2 \ , \tag{10}$$

and we find a $\chi^2/$d.o.f. close to one, and $a_1 = 4.52 \pm 0.04$, $a_2 = 2\sigma = 0.0041 \pm 0.0002$. A very good fit, and a value of the surface energy very close to the best known value of $0.0025$ from ref. [17]. We also try a power law slowing down fit, for

$$\log \tau = a_1 + a_2 \log L \ , \tag{11}$$

and we get an unreasonable $\chi^2/$d.o.f $= 8$. Things do not work so well for lower values of $T$. We are still seeing the correct exponential slowing down, but we number we estimate for the surface tension does not coincide with the number estimated in the literature. We believe the main reason for this mismatch is the poor estimation of the tunneling time due to the little number of flip-flops between the two minima and consequently the great error in the inverse of this number of flip-flops, *i.e.* the tunneling time. Also we should notice that we are on a symmetric cubic lattice, while an elongated lattice would be needed for a fair and accurate estimate.

---

[3]In this context we will not discuss about cluster algorithms, since they use the a priori knowledge of the $Z_2$ symmetry relating the two broken states to allow a (trivial) flip from the plus to the minus state. In the context of the disordered systems, where there exist many equilibrium states non related by symmetry operations, this feature is not interesting.

We have run the tempering scheme in lattice sizes $L = 10, 16, 20, 22$ and $24$ by allowing values of $\beta$ which do not lead the system in the warm phase. The system is confined in the cold phase, and we do not expect to make the tunneling from one phase to the other one very easier. This is indeed what happens. We simulate the tempering algorithm with five temperatures with $\beta_{\max} = 0.225$ fixed and $\beta_{\min}$ depends on the lattice size, for instance $\beta_{\min}(L = 24) = 0.223$, and we only analyze the data correspond to the lower temperature ($\beta = 0.225$); the best fit is by far the one to an exponential slowing down (10), with parameters $a_1 = 3.44(5)$, $a_2 = 0.0097(2)$ with $\chi^2/$d.o.f. $= 0.67$. In next section we will discuss how does the effective implementation of the tempering method performs.

## 5 Tempering Among Cold and Hot Phase

As a last step we have done what has to be done. We have set up a tempered simulation where the system is allowed to flip in and out the cold phase. We same way of reasoning we have used before to show that tempering cannot eliminate the power law critical slowing down at $T_c$ tells us now that on the contrary we can expect tempering to transform the exponential flip-flop slowing down in a power law behavior. Indeed here the number of $\beta$ steps needed to connect warm and cold phase only increases with a power law (this is a somehow an unaivodable feature of a method which improves the sampling scheme). The point is that after we have flipped to the hot phase we loose memory of the state we are coming from, and when going back in the cold phase we will fall in a random state. So, since the way in which the allowed $\beta$ windows around $\beta_c$ shrinks is dictated by a power law we expect to only find a residual power law slowing down.

The results for the pure heat bath algorithm are the ones we have discussed in the previous section. We get there an exponential slowing down basically governed by the surface tension.

In the tempered heat bath scheme we simulate lattices $L = 10, 16, 20$ and $24$ and we have allowed five temperatures with $\beta_{\max} = 0.223$ fixed and $\beta_{\min}$, again, depends on the lattice size, *e.g.* $\beta_{\min}(L = 24) = 0.2212$. We first select the points at the lower $\beta$ value (in order to be able to compare to the heat bath results). Again we try the fits to the functional forms (10), and (11). In this case the fit to an exponential slowing down is not a good fit. The $\chi^2$/d.o.f. is close to 15, more than 50 times larger than the best fit to a power dependence. The power we find in our best fit ($a_2$) is $1.76 \pm .06$, *i.e.*

$$\tau \simeq L^{1.76} . \tag{12}$$

## 6 Conclusions

Our numerical experiments have shown us that the tempering method does transform an exponential slowing down in a power law effect. Even if we have discussed the tunneling in the cold phase of the $3d$ Ising model, where these states are related by a $Z_2$ symmetry, our conclusions do not depend on such a symmetry. The fact that tempering works remarkably well for speeding up $3d$ spin glass simulations in the cold phase [6, 7] is connected to the behavior we have discussed here.

## Acknowledgments

We thank Sergio Caracciolo for an interesting discussion, and for communicating to us the unpublished results of [18]. We also acknowledge relevant communications with Margarita García and Ferdinando Gliozzi. JJRL is supported by a MEC grant (Spain). LAF and JJRL acknowledge CICyT (AEN93-776) for partial financial support.